\documentclass[reprint,showpacs,preprintnumbers,amsmath,amssymb,aps,pra]{revtex4-1}
\usepackage[utf8]{inputenc}
\usepackage{mathrsfs}
\usepackage{amsmath}
\usepackage{bm}
\usepackage{amsfonts}
\usepackage{amssymb}
\usepackage{amsthm}
\usepackage{color}
\usepackage{graphicx}
\graphicspath{{./images/}}
\usepackage{geometry}
\usepackage{hyperref}
\usepackage{ulem}
\usepackage{epstopdf}
\usepackage{graphicx}
\hypersetup{
     unicode=false,
     pdftoolbar=true,
     pdfmenubar=true,
     pdffitwindow=false,     
     pdfstartview={FitH},    
     pdftitle={My title},    
     pdfauthor={Author},     
     pdfsubject={Subject},   
     pdfcreator={Creator},   
     pdfproducer={Producer}, 
     pdfkeywords={keyword1} {key2} {key3}, 
     pdfnewwindow=true,      
     colorlinks=true,       
     linkcolor=blue,          
     citecolor=blue,        
     filecolor=magenta,      
     urlcolor=blue           
}
\usepackage[toc,page]{appendix}

\begin{document}
\title{Magnetic skyrmion generation by reflective spin-wave focusing}
\author{Xianglong Yao$^{1}$}
\author{Zhenyu Wang$^{1}$}
\email[Corresponding author: ]{zhenyuw@uestc.edu.cn}
\author{Menghua Deng$^{2}$}
\author{Z.-X. Li$^{1}$}
\author{Zhizhi Zhang$^{1}$}
\author{Yunshan Cao$^{1}$}
\author{Peng Yan$^{1}$}
\email[Corresponding author: ]{yan@uestc.edu.cn}

\affiliation{$^{1}$School of Electronic Science and Engineering and State Key Laboratory of Electronic Thin Films and Integrated Devices, University of Electronic Science and Technology of China, Chengdu 610054, China}
\affiliation{$^{2}$School of Physics and Electronics, Hunan University, Changsha 410082, China}

\begin{abstract}
We propose a method to generate magnetic skyrmions by focusing spin waves totally reflected by a curved film edge. Based on the principle of identical magnonic path length, we derive the edge contour that is parabolic and frequency-independent. Micromagnetic simulations are performed to verify our theoretical design. It is found that under proper conditions, magnetic droplet first emerges near the focal point where the spin-wave intensity has been significantly enhanced, and then converts to magnetic skyrmion accompanied by a change of the topological charge. The phase diagram about the amplitude and frequency of the driving field for skyrmion generation is obtained. Our finding would be helpful for the designment of spintronic devices combing the advantage of skyrmionics and magnonics.
\end{abstract}

\maketitle
\section{Introduction}\label{sec1}
Magnetic skyrmions are topologically protected spin textures with a high thermal stability. They normally exit in chiral bulk magnets or magnetic thin films with broken inversion symmetry, which activates the Dzyaloshinskii-Moriya interaction (DMI) \cite{Dzyaloshinsky1958,Moriya1960}. In contrast to skyrmions, magnons are the low-energy excitations in ordered magnets and can be easily generated and destroyed due to their bosonic nature. Both skyrmions and magnons have been extensively investigated and applied in information transmission and procession, which gives birth to two emerging subfields of magnetism, skyrmionics \cite{Nagaosa2013,Fert2013,Krause2016,Zhang2020} and magnonics\cite{Serga2010,Lenk2011,Chumak2015}.

The interaction between skyrmions and magnons has been widely studied in magnon-skyrmion scattering \cite{Iwasaki2014,Schutte2014}, magnon-driven skyrmion motion \cite{Zhang2017,Jiang2020}, skyrmion-based magnonic crystal \cite{Ma2015,Moon2016}, skyrmion-induced magnon frequency comb \cite{Wang2021}, etc.
Recently, the conversion between skyrmions and magnons is also attracting much attention. For example, the spin-wave emission is often observed during the annihilation and core switching of magnetic skyrmion \cite{Zhang201702,Zhang2015}. However, it is rather difficult to convert spin waves to skyrmions because the spin-wave energy is much lower than the barrier between the uniform ferromagnetic state and skyrmion. To create skyrmion by spin waves, the spin-wave energy should be accumulated to overcome the energy barrier, which has been realized by the combination of the geometry change and the DMI-induced effective magnetic field \cite{Liu2015} and by spin-wave focusing \cite{Wang2020}. In previous studies, the spin-wave focusing is achieved by constructing a spin-wave lens, which can be designed by a curved interface \cite{Toedt2016,Bao2020,Wang2020}, local graded-index region \cite{Whitehead2018,Vogel2020}, and metasurface \cite{Zelent2019,Grafe2020}. In these methods, the spin-wave reflection occurred at the interface would decrease the efficiency for spin-wave focusing. In this regard, one should avoid the spin-wave reflection as much as possible, intuitively.

However, it is known that the spin-wave can transmit through an interface without reflection only in some special cases \cite{Yan2011}. On the contrary, spin wave can be completely reflected under a more loose condition, such as at the magnetic$|$non-magnetic interface. One natural issue is how to accumulate all reflected spin waves. In this work, we design a curved film edge based on the principle of identical magnonic path length, which suggests that the parabolic film edge can focus all reflected spin waves independent of their frequencies. At the focal point, the spin-wave intensity can be significantly enhanced and the focal-point magnetization oscillates strongly and might even be locally reversed, which is considered as the precursor for the skyrmion formation.

\section{Analytical Model}\label{sec2}

\begin{figure}
  \centering
  \includegraphics[width=0.45\textwidth]{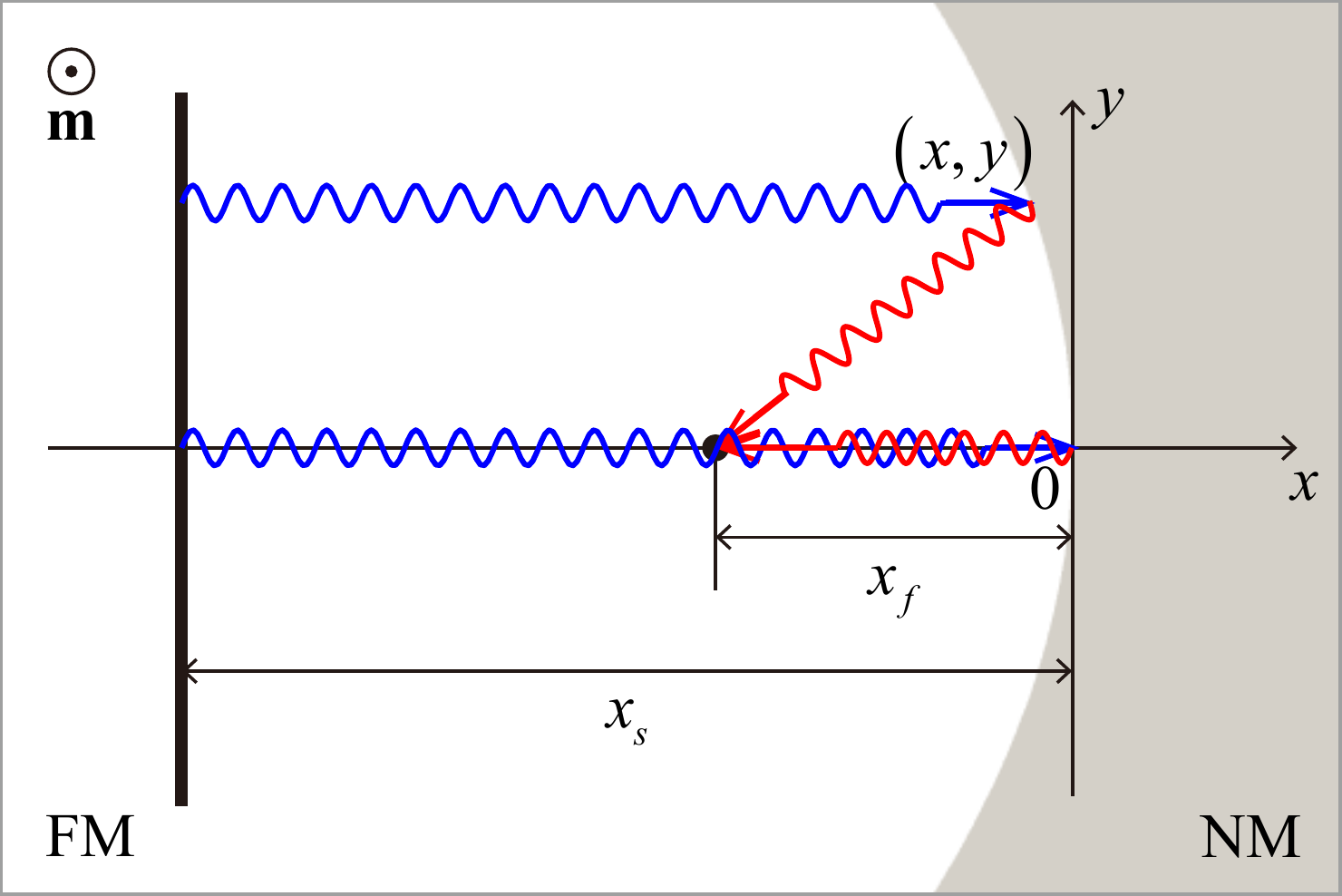}\\
  \caption{Schematic of the film edge for total-reflection focusing of spin waves. The static magnetization $\mathbf{m}$ is oriented along the $+z$ direction. FM and NM denote the ferromagnetic and non-magnetic regions, respectively. The black bar located at $x=-x_{s}$ is the spin-wave source. A parallel incident spin wave (blue wavy lines with arrow) propagates towards the curved edge and the reflected spin wave (red wavy lines with arrow) converges on the focal point at $x=-x_{f}$ (black point).}\label{fig1}
\end{figure}

We consider a chiral ferromagnetic film with a curved boundary, which is magnetized along the $+\hat{z}$ direction. The spin-wave dynamics is governed by the Landau-Lifshitz-Gilbert (LLG) equation,
\begin{equation}\label{eq_LLG}
  \frac{\partial\mathbf{m}}{\partial t}=-\gamma\mu_{0}\mathbf{m}\times\mathbf{H}_{\mathrm{eff}}+\alpha\mathbf{m}\times\frac{\partial\mathbf{m}}{\partial t},
\end{equation}
where $\mathbf{m}=\mathbf{M}/M_{s}$ is the unit magnetization vector with the saturated magnetization $M_{s}$, $\gamma$ is the gyromagnetic ratio, $\mu_{0}$ is the vacuum permeability, and $\alpha$ is the Gilbert damping constant. The effective field $\mathbf{H}_{\mathrm{eff}}$ comprises the exchange field, the DM field, the anisotropy field, and the dipolar field. In the following, the interfacial DMI is considered.

The film edge for total-reflection focusing is designed based on the identical magnonic path length (MPL) principle \cite{Wang2020}. We first consider a plane spin-wave incident from the left source ($x=-x_{s}$) which is reflected by the film edge and converges into a focal point $(-x_{f},0)$, as shown in Fig. \ref{fig1}. The identical MPL principle yields
\begin{equation}\label{eq_MPL1}
  x+x_{s}+\sqrt{(x+x_{f})^2+y^{2}}=x_{s}+x_{f},
\end{equation}
and the edge contour is described by
\begin{equation}\label{eq_edge1}
  y^{2}=-2px,
\end{equation}
where $p=2x_{f}$. One can see that the shape of the film edge is parabolic for the total-reflection focusing of the plane spin waves.

\section{Numerical results}\label{sec3}

\begin{figure}
  \centering
  \includegraphics[width=0.48\textwidth]{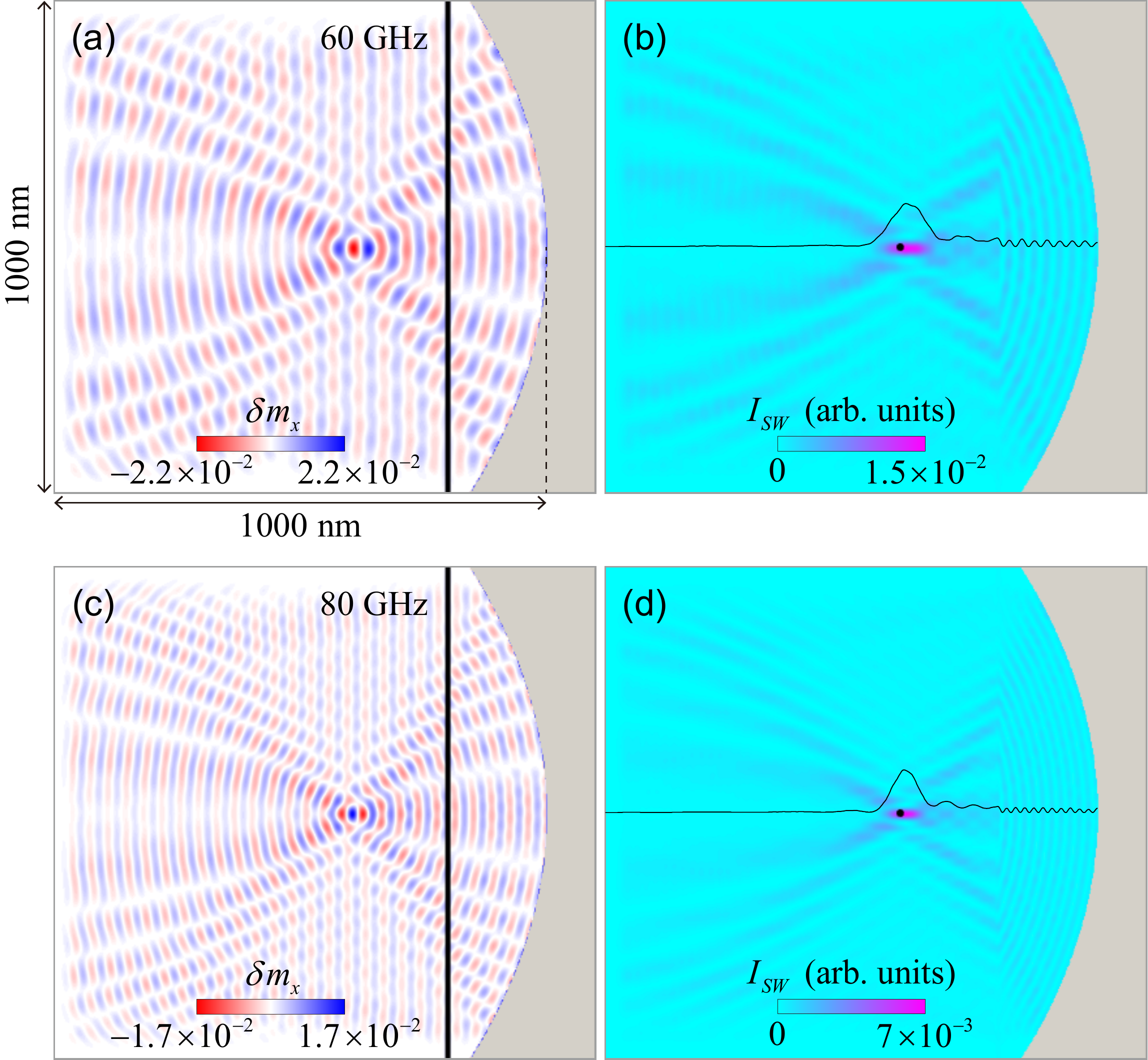}\\
  \caption{(a) Snapshot of the spin waves reflected from the parabolic edge. The spin-wave frequency is $\omega/2\pi=60$ GHz. The black bar in (a) denotes the spin-wave source located at $x=-200$ nm, which is excited by a microwave field with $\mu_{0}h_{0}=10$ mT. (b) The spin-wave intensity in (a). The black point represents the ideal position of the focal point. The black curve shows the profile of the spin-wave intensity along the $x$ axis at $y=0$. (c) and (d) show the snapshot of total-reflection focusing of spin waves and the corresponding intensity for $\omega/2\pi=80$ GHz.}\label{fig2}
\end{figure}

To verify out theoretical design, we solve numerically the full LLG equation (\ref{eq_LLG}) using the micromagnetic simulation codes MuMax3 \cite{Vansteenkiste2014}. Magnetic parameters of Co are adopted: $M_{s}=5.8\times10^{5}$ $\mathrm{A/m}$, $A_{ex}=15$ $\mathrm{pJ/m}$, $D=2.5$ $\mathrm{mJ/m^{2}}$, and $K_{u}=6\times10^{5}$ $\mathrm{J/m^{3}}$. The cell size $2\times2\times1$ $\mathrm{nm^{3}}$ is used to discretize the film in simulations. Gilbert damping constant $\alpha=10^{-3}$ is used to ensure a long-distance propagation of spin waves, and absorbing boundary conditions are adopted to avoid the spin-wave reflection by the film edges except for the curved edge \cite{Venkat2018}.

\begin{figure}
  \centering
  \includegraphics[width=0.47\textwidth]{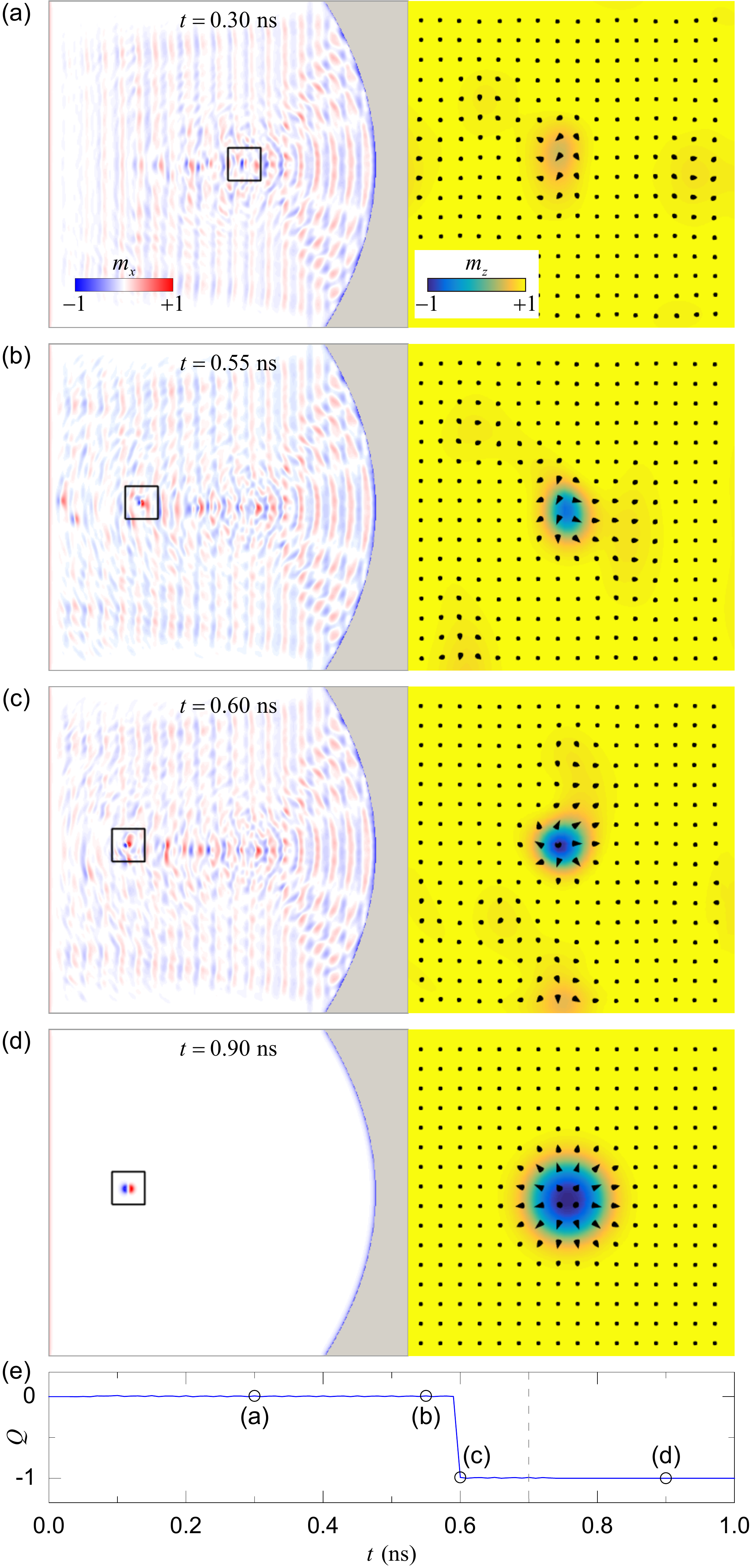}\\
  \caption{The creation process of the magnetic skyrmion induced by the total-reflection focusing. The exciting field with $\mu_{0}h_{0}=360$ mT is applied in (a)-(c) and is turned off in (d). The z-component magnetization of the rectangular area in the left column is enlarged in the right column. (e) Temporal evolution of the topological number $Q$. The microwave field starts at $t=0$ and ends at $t=0.7$ ns indicated by the gray dashed line.}\label{fig3}
\end{figure}

We first set the focal length as $x_{f}=400$ nm in simulations and design a parabolic edge to focus the reflected spin waves.
A sinusoidal monochromatic microwave field $\mathbf{H}_{\mathrm{ext}}=h_{0}\sin(\omega t)\hat{x}$ is applied in a narrow rectangular area [black bars in Fig. \ref{fig2}(a)] to excite the incident plane spin waves. Numerical results for focusing spin waves with 60 GHz are shown in Fig. \ref{fig2}(a). Using the equation $I_{SW}(x,y)=\int_{0}^{t}[\delta m_{x}(x,y,t)]^{2}dt$, we calculate the spin-wave intensity, as plotted in Fig. \ref{fig2}(b). The profiles of the spin-wave intensity is also shown [see black curves in Fig. \ref{fig2}(b)]. One can see that spin waves are reflected from the film edge and focused, leading to a significantly enhanced intensity around the focal point. However, it is found that the focal point obtained from the numerical simulation is a little shifted along $+x$ direction from the theoretical position of the focal point [black point shown in Fig. \ref{fig2}(b)]. It may be attributed to two reasons: One is the ray optic approximation for analyzing the spin-wave propagation, which requires the spin-wave wavelength (about tens of nanometers) much smaller than the size of the film edge. The other one is the spin canting at the curved edge [see Fig. \ref{fig3}(d)], which would influence the propagation of the reflected spin waves. Equation (\ref{eq_edge1}) suggests that the shape of the parabolic edge for the total-reflection focusing is independent of the spin-wave frequency. Figures \ref{fig2}(c) and \ref{fig2}(d) indeed confirm this result, which would promote the applications of the spin-wave focusing in magnonic devices.

To generate magnetic skyrmions, we increase the amplitude of the microwave field to $\mu_{0}h_{0}=360$ mT. The spin-wave intensity around the focal point is enhanced significantly, which shows a strong magnetization oscillation, as plotted in Fig. \ref{fig3}(a). With the continuous excitation of spin waves, more energies are harvested, leading to the local switching of the magnetization and the formation of magnetic droplet, which can be easily driven by spin waves [see Fig. \ref{fig3}(b)]. The magnetic droplet is a non-topological localized spin-wave soliton and is unstable in a chiral ferromagnetic film because of the high DMI energy. Under the disturbance of spin waves, the magnetic droplet is converted to a dynamical skyrmion at $t=0.6$ ns, as shown in Fig. \ref{fig3}(c). Then, we turn off the microwave field at $t=0.7$ ns and the system is relaxed toward an equilibrium state with a stable skyrmion state [see Fig. \ref{fig3}(d)].

The topological charge, which is given by
\begin{equation}\label{eq_Q}
  Q=\frac{1}{4\pi}\iint\mathbf{m}\cdot(\frac{\partial\mathbf{m}}{\partial x}\times\frac{\partial\mathbf{m}}{\partial y})dxdy,
\end{equation}
can be used to characterize the topology of skyrmions in two-dimensional systems. However, the large spatial variations of $\mathbf{m}$ in the process of the skyrmion nucleation and annihilation reduce the accuracy of the finite-difference approximation of Eq. (\ref{eq_Q}) and result in non-integer value of $Q$ \cite{Wang2020}. Recently, a lattice-based approach for computing $Q$, which does not rely on spatial derivation, was proposed \cite{Kim2020}. The topological charge in this scheme is given by the sum over the ensemble of elementary signed triangles $q_{ijk}$ on the unit sphere,
\begin{equation}\label{eq_Q2}
  Q=\frac{1}{4\pi}\sum\limits_{\langle ijk\rangle}q_{ijk},
\end{equation}
where
\begin{equation}\label{eq_qijk}
  \tan(\frac{q_{ijk}}{2})=\frac{\mathbf{m}_{i}\cdot(\mathbf{m}_{j}\times\mathbf{m}_{k})}{1+\mathbf{m}_{i}\cdot\mathbf{m}_{j}+\mathbf{m}_{i}\cdot\mathbf{m}_{k}+\mathbf{m}_{j}\cdot\mathbf{m}_{k}},
\end{equation}
which is invariant under a cyclic permutation of the indices $ijk$.
Based on the lattice-based approach, we calculate the time evolution of the topological charge $Q$ in the process of the skyrmion generation [see Fig. \ref{fig3}(e)], where nonphysical values of $Q$ are excluded. An abrupt change in $Q$ from 0 to -1 is observed at $t=0.6$ ns, which confirms the skyrmion creation from another aspect.

\begin{figure}
  \centering
  \includegraphics[width=0.5\textwidth]{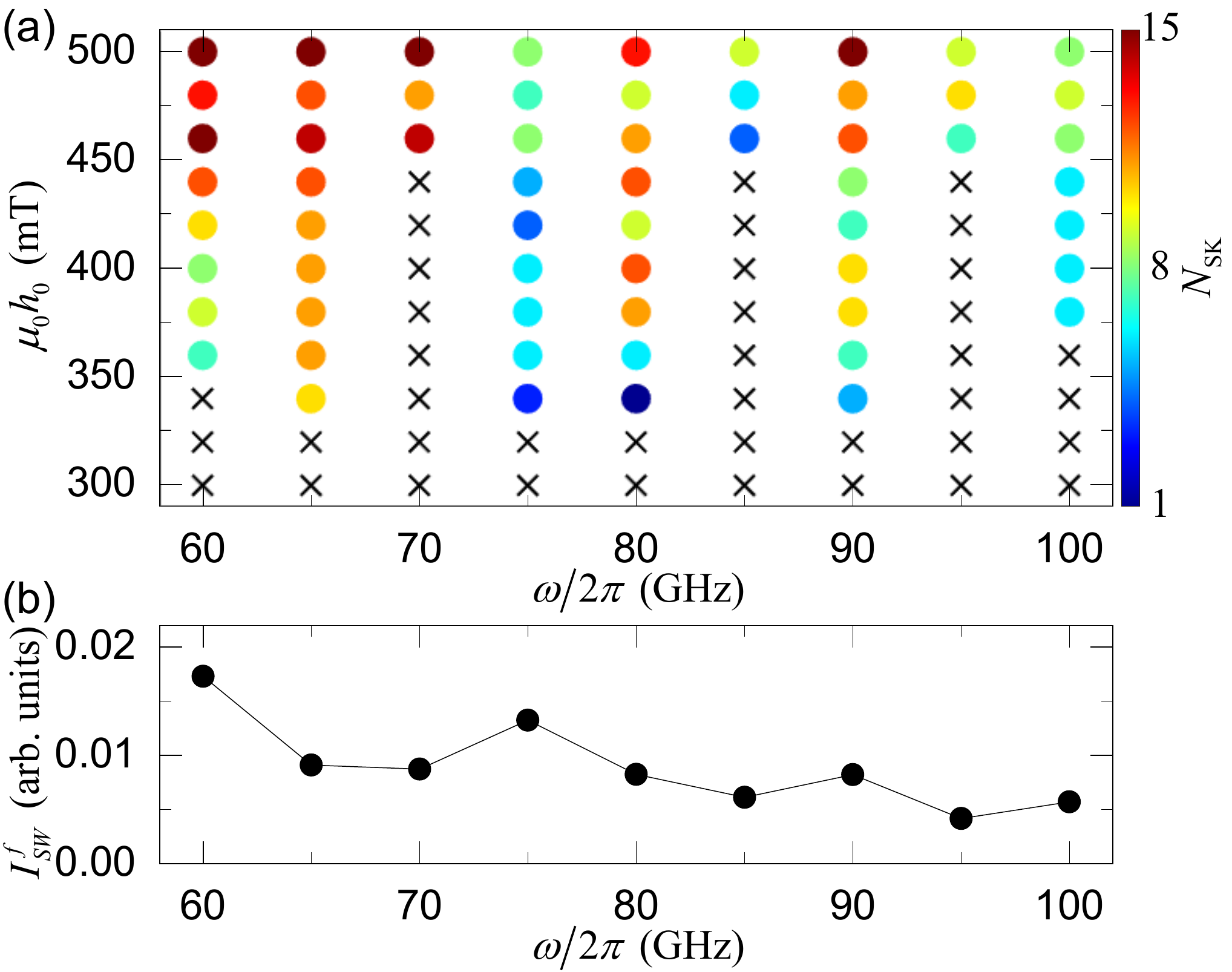}\\
  \caption{(a) Phase diagram of skyrmion generation with respect to the amplitude and frequency of the exciting field with the duration time 10 ns. The black crosses represent no skyrmion creation, the color dots denote the number of generated skyrmions. (b) The frequency dependence of the spin-wave intensity at the focal point under the microwave field with $\mu_{0}h_{0}=10$ mT.}\label{fig4}
\end{figure}

From the generation process of the skyrmion, one can see that the magnetic droplet is an indispensable intermediate between ferromagnetic and skyrmion states. Although the energy of the magnetic droplet is higher than the skyrmion, the skyrmion can not be created directly from the ferromagnetic state. This is because the droplet is non-topological and can be transformed continuously from a ferromagnetic state. However, the continuous transformation from a ferromagnetic state to skyrmion is highly unlikely, which is due to the topological protection of the skyrmion. Compared to the droplet, a change of the topological charge is accompanied for the skyrmion creation, which requires more energy input from the external driving.

Figure \ref{fig4}(a) plots the phase diagram of skyrmion generation induced by the total-reflection focusing of spin waves. As in the spin-wave focusing for the transmitted waves \cite{Wang2020}, the skyrmion number $N_{\mathrm{Sk}}$ generated by the total-reflection focusing is not monotonically increasing with the field amplitude $h_{0}$, which is owing to the skyrmion annihilation induced by the interaction between the magnetic droplet and skyrmion. In addition, it is found that the skyrmion creation at the field frequency with $\omega/2\pi=70$, 85, and 95 GHz requires a higher amplitude of the exciting field. To find out the reason for such a frequency dependence of the skyrmion generation, we plot the spin-wave intensity at the focal point for different frequencies, as shown in Fig. \ref{fig4}(b). We find that these three frequencies correspond to the local minima of the spin-wave intensity, which indicates that the higher field amplitude is required for the skyrmion formation.

\begin{figure}
  \centering
  \includegraphics[width=0.5\textwidth]{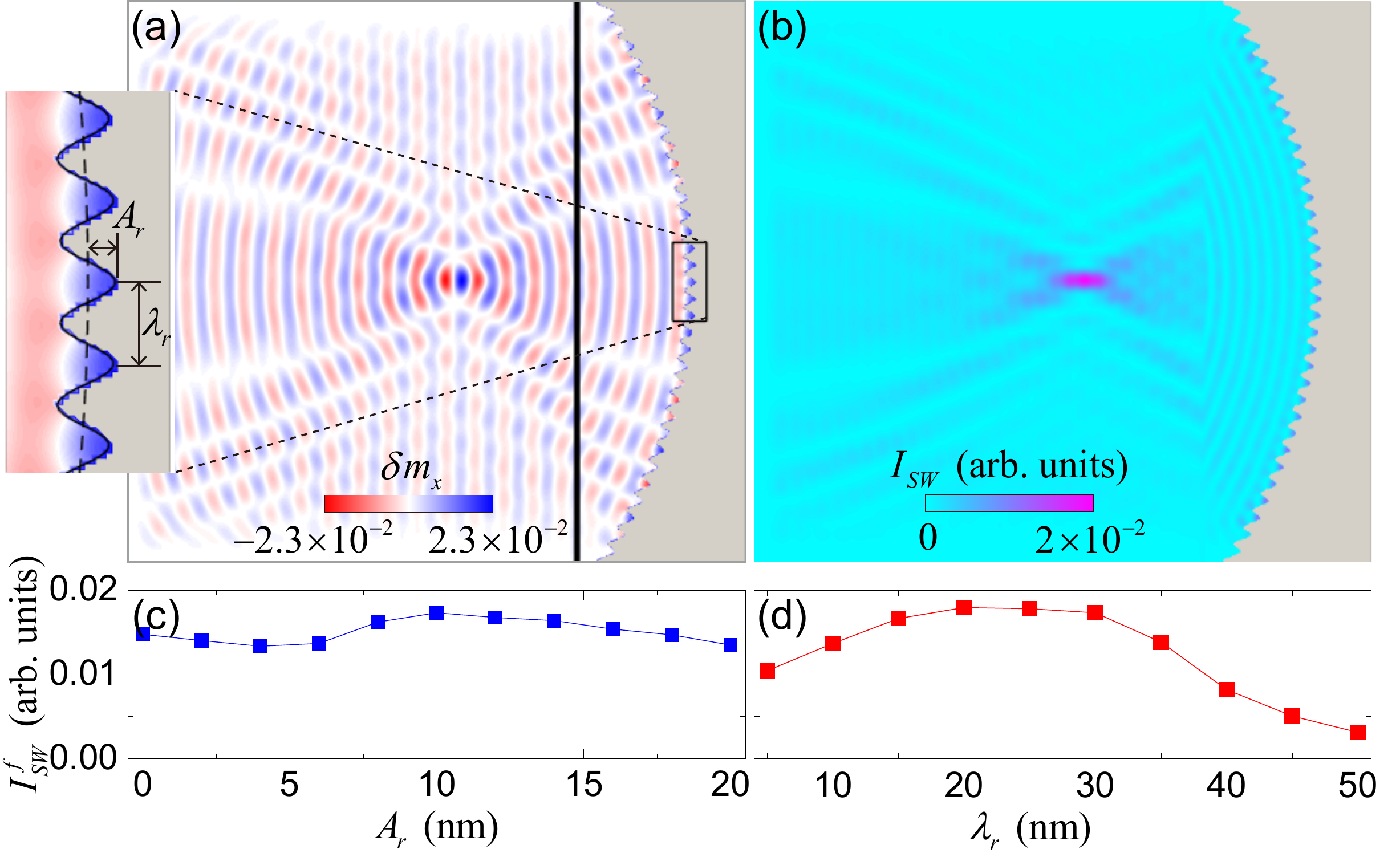}\\
  \caption{(a) Snapshot of spin waves reflected from the parabolic edge with a wave-like roughness, which is depicted by the formula $x=-y^{2}/(4x_{f})+A_{r}\cos(2\pi/\lambda_{r}y)$. The black bar shows the exciting source of spin waves with $\omega/2\pi=60$ GHz. The enlarged image of the rectangular area is shown on the left. The solid and dashed lines in the inset denote the parabolic edge with and without the roughness. The roughness amplitude and spatial period are $A_{r}=10$ nm and $\lambda_{r}=30$ nm, respectively. (b) The spin-wave intensity in (a). (c) The spin-wave intensity at the focal point $I_{SW}^{f}$ as a function of the amplitude $A_{r}$ for $\lambda_{r}=30$ nm. (d) The dependence of $I_{SW}^{f}$ on the spatial period $\lambda_{r}$ for $A_{r}=10$ nm.}\label{fig5}
\end{figure}

The above micromagnetic simulations are all conducted for the well-defined edge, which is often not the case in practical experiments, due to the roughness. To check the roughness effect on the total-reflection focusing, we performed additional simulations with a wave-like rough edge. For the rough edge with $A_{r}=10$ nm and $\lambda_{r}=30$ nm, the roughness effect can be negligible, as shown in Figs. \ref{fig5}(a) and \ref{fig5}(b). We also investigate the dependence of the roughness effect on its amplitude and spatial period. It is found that the roughness amplitude has a slight influence on the spin-wave intensify $I_{SW}^{f}$ at the focal point [see Fig. \ref{fig5}(c)], while the influence of the roughness spatial period on $I_{SW}^{f}$ is sizable [shown in Fig. \ref{fig5}(d)]. For $\lambda_{r}>40$ nm, the spin wave scattered from the rough edge is very disordered and can not be focused at the focal point (not shown here). Moreover, we note that the roughness can enhance the spin-wave intensity at the focal point under certain conditions [$A_{r}=10$ nm in Fig. \ref{fig5}(c) and $\lambda_{r}=20$ nm in Fig. \ref{fig5}(d)]. Periodic cosinusoidal roughness at the film edge has a similar profile with the parabola, which could locally focal spin waves around the edge leading to the enhancement of the spin-wave intensity.
Further investigations are necessary to elucidate the underlying physical mechanism behind such interesting phenomena. Nevertheless, the roughness effect can be ignored if its spatial period is small enough.

The results in this work are obtained in magnetic metals, which usually have high perpendicular magnetic anisotropy and high damping. A large-amplitude microwave field is needed for the skyrmion creation, which is difficult to be archived in experiments. Fortunately, magnetic insulators with perpendicular anisotropy and ultra-low damping have been demonstrated to host skyrmions \cite{Soumah2018,You2019,Caretta2020}, which makes our method more applicable from the view of materials realizations. In previous work, we propose a method to generate skyrmion by focusing the transmitted spin waves, which is realized by constructing a spin-wave lens with a curved interface \cite{Wang2020}. The shape of the interface depends on the relative refraction index of spin waves, which is frequency-dependent. Thus, that method is only feasible for focusing spin wave with one certain frequency. For spin-wave focusing with a different frequency, a new curved interface should be designed, which hinders the practical application of that method. We therefore believe that the total-reflection focusing in present study provides a promising way to generate the skyrmion.

\section{Conclusion}\label{sec4}
In summary, we investigated theoretically the skyrmion generation induced by the total-reflection focusing of spin waves. The shape of the film edge was derived based on the identical magnonic path length principle. Micromagnetic simulations were performed to confirm the focusing effect of spin waves reflected from the parabolic edge. By increasing the field amplitude, we observed the nucleation of magnetic droplet induced by the total-reflection focusing, and the transformation to the skyrmion with a change of the topological charge. Our results provide a method to generate skyrmion by reflective focusing of spin waves, which is frequency-independent, and would promote the development and application of spintronic devices combing magnons and skyrmions.

\section{Acknowledgment}\label{sec5}
We thank H. Yang and L. Song for helpful discussions.
This work is supported by the National Natural Science Foundation of China (Grants No. 12074057 and No. 11704060). Z.W. acknowledges the financial support from the China Postdoctoral Science Foundation under Grant No. 2019M653063. Z.-X.L. acknowledges the financial support of the China Postdoctoral Science Foundation (Grant No. 2019M663461) and the NSFC Grant No. 11904048. Z.Z. acknowledges financial support of the China Postdoctoral Science Foundation under Grant No. 2020M673180.


\end{document}